\documentclass[11pt]{article}
\usepackage{amsmath}
\usepackage{amsfonts}
\usepackage{graphicx}
\usepackage{hyperref}
\usepackage{url}
\usepackage{xcolor}
\usepackage{microtype}
\usepackage{booktabs}
\usepackage{fancyhdr}
\usepackage{lipsum}
\usepackage{siunitx}
\usepackage[T1]{fontenc}
\usepackage[a4paper, left=2.1cm, right=2.1cm]{geometry}

\title{Observing cities as a complex system}
\author{{Rafael Prieto-Curiel}  \footnote{Complexity Science Hub, Vienna, Austria  \hspace{200pt}  prieto-curiel@csh.ac.at}}
\date{}

\begin{document}

\maketitle

\section{Abstract} 

Cities are some of the most intricate and advanced creations of humanity. Most objects in cities are perfectly synchronised to coordinate activities such as jobs, education, transportation, entertainment, and waste management. Although each city has its own characteristics, some commonalities can be observed across most cities, such as issues related to noise, pollution, segregation, and others. Further, some of these issues might be accentuated in larger or smaller cities. For example, with more people, a city might experience more competition for space, so rents would be higher. The urban scaling theory provides a framework for analysing cities in terms of their size. New data for analysing urban scaling theory allow for an understanding of how urban metrics change with population size, whether they apply across most regions, or whether patterns correspond only to some countries or regions. Yet, reducing a city and all its complexity to a single indicator might simplify urban areas to the extent that their disparities and variations are overlooked. Often, the differences in living conditions across different parts of the same city are greater than the degree of variation observed between cities. For example, in terms of rent or crime, within-city variations might be more significant than between cities. Here, we review some urban scaling principles and explore ways to analyse variations within the same city.

\section{Introduction}

{
The world's population has grown rapidly in recent decades, and this trend will continue for decades \cite{bai2016defining}. Some countries, such as Niger or Chad, will double their 2020 population before 2050 \cite{Tusting2019}. This population growth mostly happens in cities. In fact, the rural population worldwide has already reached its peak (nearly 3.4 billion people) and will decline in the coming decades \cite{UnitedNationsDESA}. Thus, although the world's population will continue to increase, that growth will occur only in cities. Whilst many parts of the world have already undergone urbanisation, the next three decades will bring sweeping changes in many parts, including Africa and South Asia \cite{angel2020shape, seto2012global, pumain2004scaling}. Due to urbanisation and population growth, some cities will continue to grow at unprecedented speed and might reach a population of 80 million or more \cite{100MillionCities}. How could a city with 80 million inhabitants function? What shape will it have, or what surface will it occupy? How long will people spend on their daily commute?
}

{
Cities offer unparalleled access to a wide array of essential services, including healthcare, education, transportation, cultural amenities, water, and sanitation services \cite{Leon08, Combes12, Glaeser01, Winters11}. For example, in 2022, 81\% of the world's urban population had access to safely managed drinking water services, but only 62\% of the rural population did \cite{JMPDatabase}. Similarly, 65\% of the world's urban population had access to safely managed sanitation services, against 46\% of the world's rural population. The difference between urban and rural populations is even more marked in poorer parts of the world. Cities are hubs where service integration enhances quality of life and where mass access to services is achieved. Between 2000 and 2022, 1.4 billion people obtained access to safely managed water in urban areas, whilst only 0.7 billion people gained access in rural areas \cite{JMPDatabase}. Although cities are where most people live and obtain essential services, they also pose diverse challenges, including fierce competition for resources such as water \cite{wight2021mapping, florke2018water}. Cities create huge issues such as pollution, damage to the ecosystem, loss of biodiversity, and land-cover change, whilst some social aspects are also challenging in cities, such as extensive commuting times, crime, violence, and social disparities \cite{Seto12, SetoReenberg12, Gong12}. 
}

{
This chapter describes how cities may be analysed quantitatively in many ways. Firstly, by comparing cities, it is possible to detect if there is a pattern among them. For example, whether a city has more restaurants than other cities or whether it receives more tourists. Thus, the first step is to construct an urban indicator (say, the number of restaurants). Secondly, by correlating those patterns with some covariates. For example, we could analyse whether a city has more restaurants if it is a coastal city or a tourist destination, so the covariates considered could be the distance to the nearest beach or the number of tourists in a year. The most common and perhaps most significant covariate to analyse is city size, so correlations between population size and urban indicators are often considered. Finally, we analyse the same indicator at more granular levels. For example, we analyse whether restaurants in the city are uniformly distributed or whether some neighbourhoods have more restaurants. Although many spatial covariates could be considered at the city level, distance to the city centre often shows a marked pattern, with discrepancies between central parts of a city and remote locations.
}

\section{How to analyse cities using complex systems}

{
Our understanding of societies has rapidly evolved in recent decades, mainly due to new data that have opened new ways to observe patterns, often at a global scale \cite{prieto2023scaling, candipan2021residence, Tusting2019, guan2020delineating}. For example, nearly 20 years ago, an experiment was conducted to track where banknotes appeared. Tracing two successive appearances of the same bill (often appearing in distinct locations, suggesting that a person travelled from the first to the second location), relevant patterns of human travel were quantified and mapped for the first time at such a large scale \cite{ScalingHumanTravel}. Similarly, data from mobile phone calls or social media have been used to model mobility patterns, opinion dynamics, and other social patterns such as segregation or polarisation \cite{gonzalez2008understanding, bakshy2015exposure, GravitationInteraction, MobileCommunication}. This way of observing societies through the lens of large amounts of data is possible due to novel methods of capturing information, increasing processing power, and new models for analysing them. 
}

{
At the urban level, novel data and models have also contributed to understanding how we live in cities. For example, credit card data was used to understand lifestyles in urban populations \cite{di2018sequences}. Also, social media and real-time GIS data enabled mobility forecasting for traffic control, improved public transport systems, and enabled the design of ``smart'' cities \cite{BattyGIS, HumanSensing}. Percolation models and real-time high-resolution GPS data have improved our understanding of real traffic jams \cite{zeng2019switch, zhang2019scale}. Similarly, collective pedestrian movement has been analysed as a self-organising process \cite{helbing1995social, SelfOrganisedPedestrians}. These types of analysis require large amounts of data and mathematical models based on complex systems. Thus, this way of observing cities is only possible in the past few years. 
}

{
New data and models have also substantially changed how we observe infrastructure in cities. One interesting dataset comes from a collaborative project called OpenStreetMap \cite{OpenStreetMap}. The project aims to create a free, editable map of the world, built by a community of mappers who contribute data about roads, railway stations, and more. The data from OpenStreetMap is freely available and can be used by anyone for various purposes, including navigation, research, and application development. Given its collaborative nature, it has many benefits, particularly because users maintain it, creating an open-access dataset. However, it also has some negative aspects, particularly that not all roads or infrastructure have been mapped and that nobody can confirm whether the infrastructure is still there after it has been added \cite{prieto2022constructing}.
}

{
Beyond road infrastructure, it is now possible to analyse the locations and footprints of all buildings in many regions of the world \cite{sirko2021continental, milojevic2023eubucco}. Obtaining data for millions of buildings requires AI tools, high-resolution satellite imagery, internet bandwidth to share massive amounts of data, and computing power to process more than 1.8 billion building detections. For many limitations, this was an impossible task just a few years ago. Further, building data can be combined with other novel sources, such as precipitation, solar radiation, wind speed, and other climatic variables and terrain attributes to model correlations between human settlements and weather \cite{fick2017worldclim, sirko2021continental, esch2022world, NASAJPL2020}. 
}

{
It is possible to analyse the structure of a city using buildings' footprint data, considering its visible infrastructure (as underground infrastructure is not detected) \index{Buildings!footprint}. This type of data usually does not include roads or transport infrastructure, such as railways. Instead, it contains the polygons that form the footprints of each building (see, for example, all buildings in Basel in Figure \ref{BaselFig}, produced using individual building data \cite{milojevic2023eubucco}). Some interesting calculations are possible, such as the orientation of some neighbourhoods or the building size distribution. The data opens new ways to analyse cities based on what is constructed.   

\begin{figure}[!htbp]
\centering
\includegraphics[width=0.6\textwidth]{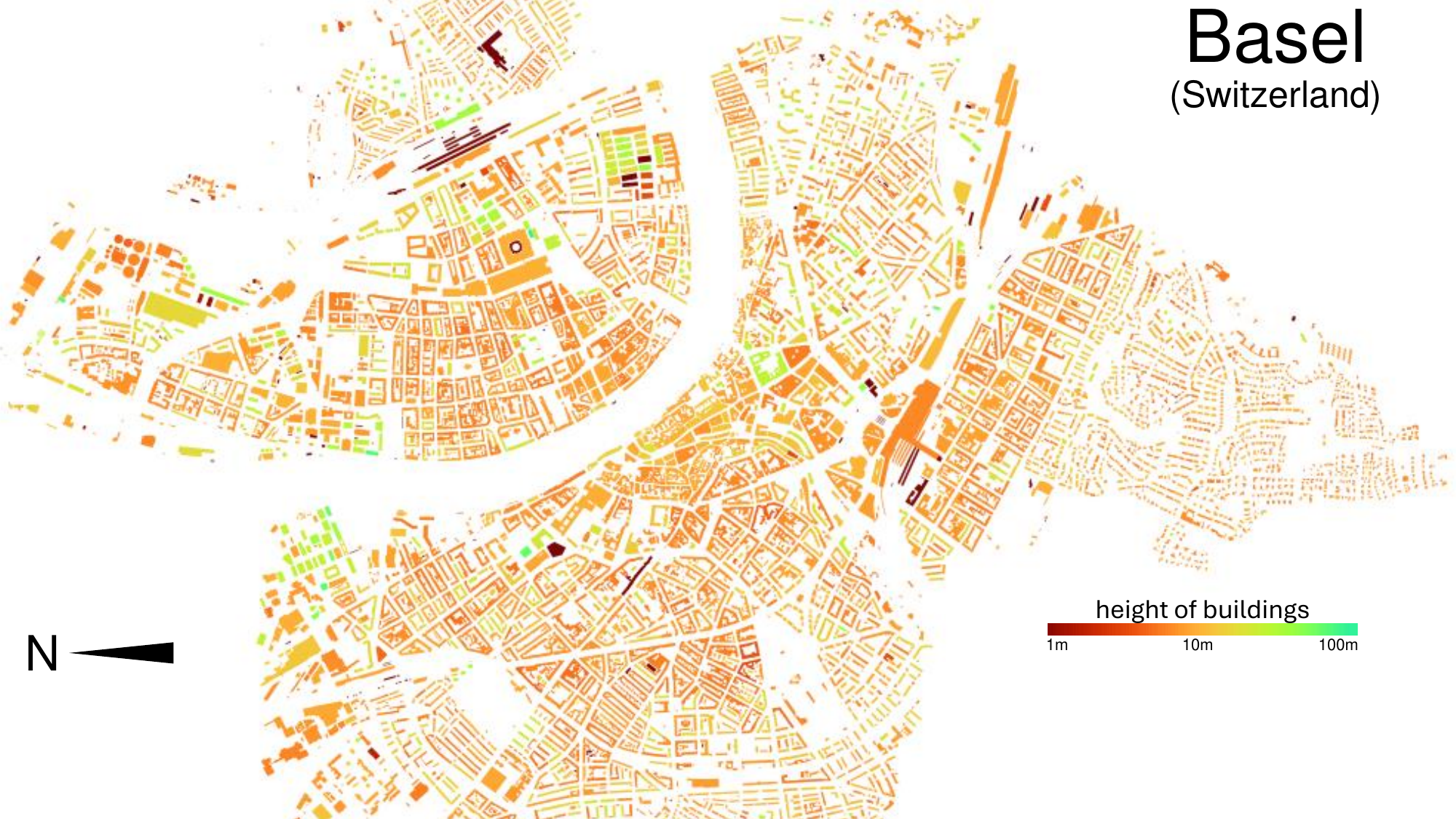}
\caption{Buildings detected in Basel, Switzerland. The colour corresponds to the (log) height of buildings. Data from \cite{milojevic2023eubucco}.}\label{BaselFig}
\end{figure}
}

{
Building data opens new ways to observe cities and the complex systems they form. For example, when you walk through a neighbourhood, the buildings and houses often seem to be of similar height. Perhaps it is reasonable to think that if most of the structures in a neighbourhood are two to three stories tall, adding a much taller building would stand out and disrupt the area's visual flow. Conversely, if a neighbourhood consists mainly of high-rise buildings, a single-story house may look out of place. Although many factors may explain why this happens (such as zoning regulations, architectural styles, historical development, economic factors, and more), it is relevant to first ask whether this is the case. Are neighbourhoods formed mostly of buildings of similar height?
}

{
One way to answer this question is by using building data. Consider, for example, all buildings constructed in Vichy, France, and their heights (Figure \ref{VichyFig}). For building $j$, we express its height in metres as $h_j$ and its age in years as $g_j$. Then, for building $j$, we can find which one is closest to it, say $k$, and compute the difference between their heights $d_j = |h_j-h_k|$. If we compute this metric for all buildings, we can analyse its distribution and compute metrics like the average. 
\begin{figure}[!htbp]
\centering
\includegraphics[width=0.6\textwidth]{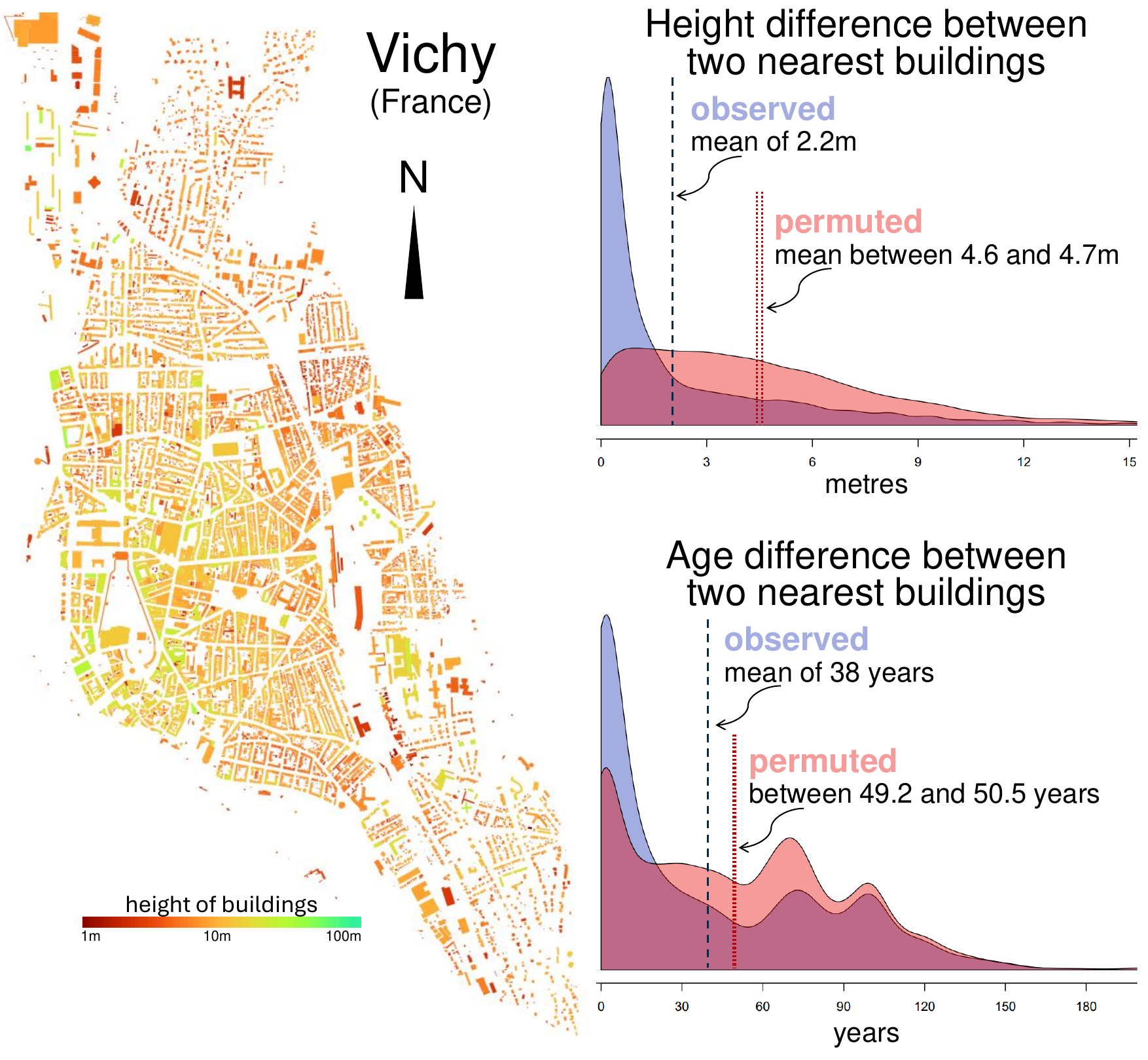}
\caption{Buildings in Vichy, France. Vichy is a spa town in the centre of France with roughly 25,000 inhabitants and with many centuries of history. The colour corresponds to the (log) height of buildings. Data from \cite{milojevic2023eubucco}. The right panels correspond to the distribution of the height difference between any pair of nearest buildings (top) and the age difference between them (bottom). The blue distribution corresponds to the observed display of a city, and the red distribution corresponds to a permuted city, where we re-arrange the height (or age) of all buildings in a random location. }\label{VichyFig}
\end{figure}
}

{
In Vichy, the two nearest buildings have an average height difference of 2.2m. Although some pairs of buildings have a much larger difference, we observe that most nearby buildings are similar in height. However, is 2.2m a small difference? Could this be the result of a random distribution of building height in the city? We test this hypothesis as follows. We take all buildings in the city and randomly permute their heights. That is, the city has the same layout, but buildings are assigned heights relative to one another within it. Then, we measure the height difference between the nearest buildings, say $d_j^P$, analyse their distribution, and compute the average. If the distribution of $d_j$ and the permuted city $d_j^P$ is similar, then the observed height of buildings might be observed simply by randomness. However, this is not the case. In Vichy, the permuted city has, on average, more than double the height difference between the two nearest buildings. We can repeat the same permutation and obtain a similar difference between the nearest buildings. A common practice to construct confidence intervals is to repeat the same type of permutation many times (usually 1,000 times is enough), drop the top and bottom extremes, and keep the rest as what could be observed by randomness. By considering permutations of the height of buildings in the city, we obtain that, by chance, we should observe a difference of 4.6 to 4.7m between the nearest buildings, more than twice the observed difference of 2.2m (Figure \ref{VichyFig}). Thus, at least in Vichy, France, we are sure that some neighbourhoods have tall buildings and some have shorter buildings, and thus, the buildings and houses often seem to be of similar height.
}

{
We can do a similar test for the age of buildings and houses in a city. If we take the difference in age between the two nearest buildings, say $\delta_j = |g_j-g_k|$, and analyse its distribution, we can analyse if neighbourhoods have buildings of similar age. In Vichy, the average difference in age between the two nearest buildings is 38 years (Figure \ref{VichyFig}). With a permutation of the age of buildings, we obtain that a difference between 49.2 and 50.5 years should be observed. Thus, nearby buildings also have a similar age. Notice, however, that the age between the nearest buildings is much closer to the one observed by randomness than the height. Thus, even if two nearby buildings are constructed many years apart, they tend to preserve similar heights. 
}

{
Building data at the street and neighbourhood levels has opened new ways to observe cities globally. Are buildings of similar height across neighbourhoods also observed elsewhere? Particularly in more modern cities? Or in bigger cities? Yet for many metrics that can be studied at the city level, a fundamental issue must be considered: cities vary widely in size. One of the most critical elements to consider in the analysis of cities is their population. The number of inhabitants of urban areas might vary from a few thousand people to a large metropolis of millions of people. Thus, it is relevant to understand how cities change depending on their size. For example, we could simply consider the number of buildings as an urban metric. But that number is closely related to the population of that city. Cities with thousands of people have thousands of buildings, whereas cities with millions of people also have millions of buildings. Thus, it is worth constructing urban indicators that do not depend on the population in an obvious manner.
}

\section{Urban scaling}

\begin{figure}[!htbp]
\centering
\includegraphics[width=0.6\textwidth]{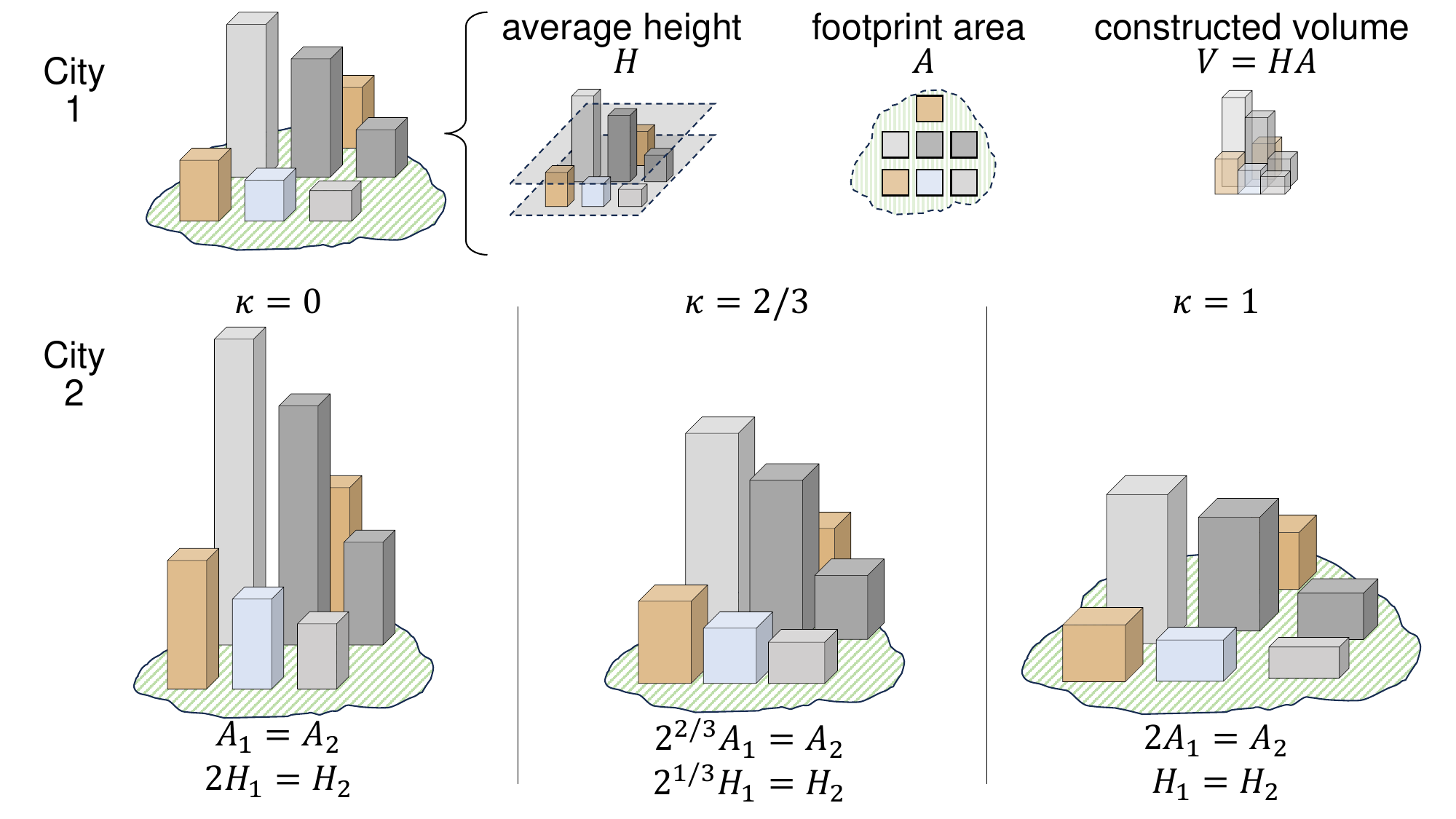}
\caption{It is possible to capture the average height of buildings in a city, as well as the constructed surface and volume. For a city with twice the population (bottom), we may assume that it also has twice the constructed volume. Yet, there are many configurations for how the city may expand its built volume, either by increasing height, surface area, or both. }\label{ScaleArea}
\end{figure}

{
In many ways, a large city is not just a scaled version of a small city. Most components and aspects of a city vary with its size. And this variation might not scale linearly with the population. For example, take two cities, $C_1$ and $C_2$, with a population of $P_1$ of one million people and $P_2$ of two million people (so $P_2 / P_1 = 2$). Then, consider its constructed volume, $V_1$ and $V_2$. Should it follow that $V_2 / V_1 = 2$ as well? This scenario would be the case if every person requires a fixed amount of constructed infrastructure, say $\mu$ m\textsuperscript{3}. Then, what about the constructed surface in cities? Consider the surfaces $S_1$ and $S_2$. Should it follow that $S_2 / S_1 = 2$ as well? If large cities are not simply a scaled version of small cities, then we can consider three scenarios (Figure \ref{ScaleArea}). One scenario is when a city only has vertical growth. In that case, we would obtain that $S_2/S_1 = 1$. A second scenario is where the growth is only horizontal, so we put a city like $C_1$ next to a copy of itself to make city $C_2$. In such a case, $S_2 / S_1 = 2$. However, a third scenario is when both height and surface grow at similar rates, so a large city is a scaled version of a small city. In such a case, it is possible to show that $S_2 / S_1 = 2^{2/3}$. 
}

{
The growth of the city surface as it doubles its population can be expressed by a factor of $2^\kappa$, where $\kappa = 0$ indicates vertical growth, $\kappa = 1$ is a horizontal expansion, and $\kappa = 2/3$ corresponds to a scaled growth in all dimensions. Also, values above $\kappa = 1$ suggest not only horizontal growth but also that the city adds even more infrastructure with more population (for example, an airport, a stadium, or an industrial park). In reality, most scenarios are reasonable and depend too much on the local conditions and the budget. In Hong Kong, for example, the population nearly doubled between 1970 and 2020, but in a region where the space is severely constrained, most of its growth was vertical (so $\kappa_{HK} \approx 0$). The contrary has happened in some cities in Africa, where most of its growth has been horizontal (so $\kappa_{Af} \approx 1$) \cite{prieto2023scaling}. Analysing and comparing cities involves constructing indicators that capture the main elements of these different configurations.
}

{
In general, urban scaling studies aim to capture how population variation affects aspects such as surface area, infrastructure, and social life \index{Urban!scaling}. Such studies have shown, for example, that people from larger cities tend to produce more patents, have a higher income per capita, and some types of diseases, but are also prone to more congestion, road accidents, and some diseases \cite{keuschnigg2019urban, cabrera2020uncovering, louf2014congestion, zhou2020gini}. Smaller cities usually require more infrastructure per person, such as the road surface or the petrol stations \cite{GrowthBettencourt}. People from larger cities migrate less and are more likely to return after moving \cite{ScalingMigrationRPC}. 
}

\subsection{Quantifying urban scaling}

{
One mathematical expression that helps capture the impact of city size is the equation
\begin{equation} \label{scalingEq}
Y_i = \alpha P_i^\beta,
\end{equation}
where $Y_i$ is the variable of interest for city $i$ and $\alpha$ and $\beta$ are unknown parameters. For example, $Y_i$ could be the number of restaurants in the city $i$. We obtain the coefficients $\alpha$ and $\beta$, usually through a regression. With $\beta>1$ the variable of interest is called ``superlinear'' and indicates that large cities have higher values of the variable $Y$ per capita (since the per capita rate $Y_i/P_i$ is given by $\alpha P_i^{\beta-1}$). With $\beta<1$, results are ``sublinear'', and with $\beta \approx 1$, city size has little or no impact on the per capita rate of that city. Therefore, the coefficient of interest is usually $\beta$ and values above or below $\beta = 1$ are critical. When using equation \ref{scalingEq}, we are not assuming a linear correlation with the population, although it can also be a result if values of $\beta$ are close to one. 
}

{
For some social indicators in the USA, $\beta = 1.15$ was obtained. The coefficient means that if we compare two cities, $C_1$ and $C_2$, where the population of $C_2$ is two times the population of $C_1$, the expected values of $Y_2$ are approximately $2^{1.15} = 2.22$ times larger. On a per capita basis, in the city $C_2$, there are 11\% more of $Y$ than in city $C_1$. Values of $\beta \in (0, 1)$ indicate that the corresponding variable $Y$ increases with city size but at a slower rate than population (so the per capita ratio decreases with size). Finally, the same mechanism can be used to detect whether a city indicator decreases with population size (for example, green areas or wildlife) when $\beta < 0$.
}

\subsection{Why do we observe urban scaling?}

{
Before trying to understand why we observe urban scaling, it is relevant to highlight that for many scaling properties, it remains unclear whether the coefficients observed in one country can be exported to other parts of the world or whether some of the results depend on how cities are defined or what is being measured \cite{arcaute2015constructing, louf2014scaling}.
}

{
Although some aspects of urban scaling remain unclear, some explanations for why we observe them are interesting. Some of the urban scaling properties are usually attributed to economies of scale, based on a utility maximising mechanism for some economic agents living in those cities \cite{Pumain00, pumain2004scaling, GrowthBettencourt}. Also, by drawing some similarities between the body mass and the metabolic rate of biological systems and cities and their population, some scaling related to urban infrastructure has been delineated \cite{GrowthBettencourt}. Some explanations of why we observe some scaling have been related to the resources needed to introduce a new member into the city \cite{GrowthBettencourt}, the number of interactions between different people \cite{bettencourt2013origins}, transportation costs, economies of scale \cite{bettencourt2020interpretation}, the selective migration of highly productive people \cite{keuschnigg2019urban}, the distribution of GDP \cite{ribeiro2021association}, among many other theories \cite{RIBEIRO20231}. However, it remains unclear why we observe some of the scaling properties among cities, whether they are universal and persistent, or whether they are mostly observed in urban areas in the USA. 
}

{
Urban scaling and its implications are quite relevant, particularly for some of the scaling coefficients related to social life. For example, it has frequently been argued that large cities suffer more crime than small cities \cite{SacerdoteCrimeCities, CrimeAndUrbanFlight, GrowthBettencourt}. For instance, larger Brazilian cities report more homicides per 100,000 inhabitants than smaller ones, with similar trends observed for homicides in Colombia, burglaries in Denmark, and property crime in the USA \cite{alves2013scaling, gomez2012statistics, oliveira2021more, chang2019larger}. Yet, even if we accept that crime scales superlinearly with city size, would that ``condemn'' large cities to always be insecure? Can we not construct safe metropolitan areas and big cities? Indeed, opposite indicators have also been observed, where larger cities are less violent, such as murders and homicides in India and burglary in South Africa \cite{sahasranaman2019urban, oliveira2021more}. Thus, urban scaling is an interesting angle from which to observe huge disparities in city size, but results need to be taken with caution and with an understanding of its possible implications.
}

\section{Scaling infrastructure of cities}

{
Imagine a city with $n$ buildings, each occupied by $m$ people, with the same area (say, one m\textsuperscript{2}). If buildings are arranged one next to the other to minimise the average distance between them, they would be arranged circularly and compactly, occupying a surface of $n$ m\textsuperscript{2}. For a sufficiently large number of buildings, they will form approximately a circle with an area of $n$ m\textsuperscript{2}. The mean distance between those buildings would be $128 \sqrt{n} /(45 \pi) $ \cite{prieto2023scaling}. Therefore, in a city with a circular and compact shape, distances grow with the square root of the population. With the dataset that gives the location of buildings in a city, it is possible to measure the distance between each pair of them. Thus, we can take all the buildings that correspond to some city, measure the distance between every pair of them, and then compute their average. For a set of cities, $i = 1, 2, \dots, m$ and their buildings, we can measure the average distance between the buildings for each city, $D_i$. Then, we aim to understand the correlation between the population in cities, $P_i$, and the mean distance between their buildings, $D_i$. With the equation
\begin{equation} \label{ScalingDist}
D_i = \alpha_D P_i^{\beta_D},
\end{equation}
we express the observed mean distance in cities in terms of the population of each city. There are two unknown coefficients, $\alpha_D$ and $\beta_D$, which capture the impact of population. For estimation, we can take the logarithm on both sides of equation \ref{ScalingDist} and obtain a linear expression in terms of $\log \alpha_D$ and $\beta_D$. As a baseline, we know that a coefficient of  $\beta \approx 1/2$ suggests that cities grow horizontally and follow similar patterns. A coefficient $\beta < 1/2$ suggests a more round or compact growth in terms of surface (possibly with vertical expansion). With $\beta > 1/2$, cities expand mostly in horizontal ways, increasing distances faster than the natural reasons attributable to population. By considering the distance between buildings of more than 6,000 African cities, it was obtained that the coefficient $\beta_D = 0.543$, so distances in cities grow at a faster rate than the horizontal case ($1/2$) \cite{prieto2023scaling}. 
}

{
Why are distances in cities in Africa above the threshold of $\beta_D = 1/2$? There are four explanations related to the way cities expand. Either 1) large cities have bigger buildings, 2) they grow in non-circular ways, 3) they expand in less compact manners, or 4) the actual number of buildings in cities grows faster than the population. We can decompose the surface of a city into four multiplicative components: $B_i$ is the number of buildings in the city, $A_i$ is their average area, $S_i$ is the \emph{Sprawl} (corresponding to the space between buildings, where smaller values are observed in compact cities, and $E_i$ is the \emph{Elongation} of a city, where smaller values indicate a round shape and higher values suggest a more elongated footprint. Thus, the area of city $i$ is expressed as $B_i A_i S_i E_i$ and the mean distance between two buildings is given by
\begin{equation} \label{MeanDistCircle}
D_i = \frac{128}{45\pi} \left( \underbrace{B_i A_i}_\text{footprint} \underbrace{S_i E_i}_{\text{shape}} \right)^{1/2}.
\end{equation}
The BASE model decomposes distances into four multiplicative components \cite{prieto2023scaling}. Inspired by an ellipse, we measure the Elongation of a city. The mean distance between two random points inside an ellipse has no closed solution, but an approximation can be constructed by considering the ratio between the major and the minor axis \cite{parry2000probability}. Thus, we define the Elongation $E_i$ as
\begin{equation} \label{EccEqn}
E_i = \frac{\sqrt{\pi} M_i}{2\sqrt{B_i A_i}},
\end{equation}
where $M_i$ is the longest distance between any two buildings (so the major axis of the ``ellipse'' formed by the city) and where $2\sqrt{B_i A_i / \pi}$ is the smallest possible diameter of a circle with $B_i A_i$ as a footprint (so the minor axis of that ``ellipse''). We obtain a coefficient $E_i \geq 1$ concerning the number or area of buildings, where $E_i = 1$ indicates a perfectly round city. Here, if a city is a scaled version of another, the elongation $E_i$ should remain the same. For example, in a city that has four times the number of buildings, distances also grow, including doubling the maximum distance and doubling the smallest diameter, so $E_i$ remains the same. From the data corresponding to the buildings of a city, we can directly measure the number of buildings, $B_i$, their mean area $A_i$, the mean distance $D_i$, and the maximum distance $M_i$, so we can also compute the Elongation $E_i$ and obtain the Sprawl, given by 
\begin{equation} \label{SprawlEquation}
S_i = \frac{45^2 \pi^{3/2} D_i^2}{2^{13} M_i \sqrt{B_i A_i}} = \gamma \frac{D_i^2}{M_i \sqrt{B_i A_i}},
\end{equation}
for $\gamma = 45^2 \pi^{3/2} / 2^{13} \approx 1.38$. The most relevant part is that the components of the BASE model are comparable across cities of different sizes. One way to interpret the Elongation and the Sprawl of a city is that the mean distance between buildings in cities increases proportionally to $\sqrt{S_i E_i}$, so if a city has an elongation value of $E_i = 4$, then the mean distance between buildings is double because the city is Elongated. Similarly, for the Sprawl $S_i$ and its impact on distances.
}

{
To analyse the impact of the population, we can compute the scaling coefficient of each component of the BASE model separately. First, regarding the footprint of a city ($BA$). In Africa, the number of buildings within a city grows with population, with roughly one extra building for every 2.6 people \cite{prieto2023scaling}. However, that number decreases slightly with size, reflecting some shared infrastructure, with $\beta_B = 0.981 \pm 0.007$. The average size of buildings also grows slightly with city size, with a coefficient $\beta_A = 0.067 \pm 0.005$. That means that if city $C_2$ has twice the population of the city $C_1$, then the average size of their buildings is $A_2/A_1 = 2^{0.067} \approx 1.05$. Thus, in a city with twice the population, the average building area increases by about 5\%. 
}

{
In terms of the shape of the city ($SE$), there are also some effects in terms of city size. The Sprawl and the Elongation across cities in Africa vary considerably, from compact circles to highly elongated, sprawling shapes (Figure \ref{Infograf}). Yet, we find only a minor effect of city size on the Elongation or the Sprawl of a city, meaning that as cities grow, they are not forming rounder or more compact shapes, on average.  In Africa, we observe that $\beta_E = 0.005 \pm 0.005$, meaning that cities are equally round if they are small or big cities. Also, we observe that $\beta_S = 0.011 \pm 0.006$, meaning that cities are slightly more sprawled as they grow.

\begin{figure}[!htbp]
\centering
\includegraphics[width=0.6\textwidth]{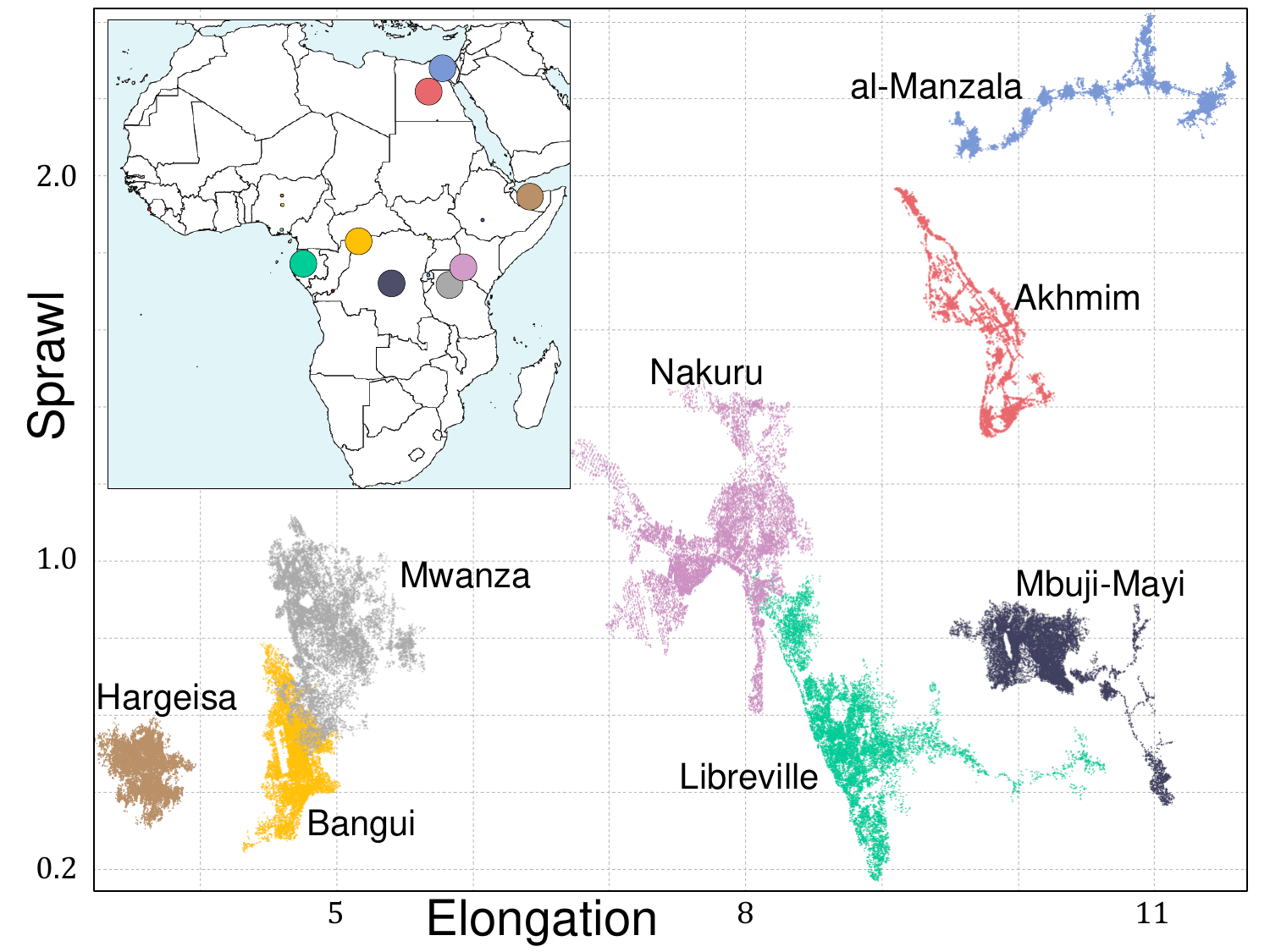}
\caption{Eight cities of similar population in Africa are plotted using the same scale. Each building in the city is represented by a tiny dot, revealing the city's internal patterns. The horizontal axis is the Elongation, and the vertical axis is the Sprawl of each city. Cities are centred on their corresponding Elongation and Sprawl. Round and compact cities are on the bottom left, whereas elongated and sprawled cities are on the top right corner. }\label{Infograf}
\end{figure}
}

{
Combining the effect of the number of Buildings, their increasing Area, and the negligible effect of the Sprawl and Elongation, we can express the distance between buildings by its four components (from equation \ref{MeanDistCircle}). Thus, the fact that in Africa, the obtained scaling coefficient for the mean distances is $\beta_D = 0.532$ is mostly because the number of buildings increases sublinearly (with $\beta_B = 0.981 \pm 0.007$), but also because the size of their buildings increases (with $\beta_A = 0.067 \pm 0.005$), their Sprawl remains almost the same (with $\beta_S = 0.011 \pm 0.006$), and their Elongation also remains almost the same (with $\beta_E = 0.005 \pm 0.005$). Combining all coefficients in Equation \ref{MeanDistCircle}, we get that 
\begin{eqnarray*}
D_i &\propto& (B_i A_i S_i E_i)^{1/2} \\
&\propto& P_i^{\frac{0.981+ 0.067 + 0.011 + 0.005}{2}} = P_i^{0.532},
\end{eqnarray*}
and thus, distances in large cities grow slightly faster than the square root of its population, with exponent $\beta_D = 0.532$. 
}

{
With the system of Equations \ref{MeanDistCircle}, \ref{EccEqn}, and \ref{SprawlEquation}, it is possible to imagine a city with a population of 80 million inhabitants. If present trends continue, it will have 18 million buildings (roughly two-thirds of the number of buildings in Nigeria today) with a footprint of 175 km\textsuperscript{2} and an average distance between buildings of over 85 km. The burden on city dwellers of such distances could force the city to grow roughly circular, compact, and more vertical.
}

{
Cities face two opposing forces as they grow. People demand more infrastructure, so more and larger buildings. However, as the population increases, distances grow and become critical, so cities experience intense competition for space and make better use of it, which results in less elongated cities \cite{batty2008size}.
}

\subsection{The Line in Saudi Arabia}

{
Saudi Arabia plans to construct a new city, called The Line, home to 9 million people. The most relevant aspect is its form, a line with a surprising length of 170 km, enough to travel from the East to the West coast of Italy \cite{prieto2023arguments}. It aims to host millions in a footprint of only 34 km$^2$. Unlike any other city, The Line is conceptually a straight line stretching from the Red Sea, 170 km East. The whole city is planned to consist of two unbroken, jointed lines of 500 m-high skyscrapers, each 200 m wide. The buildings in The Line will be taller than the Empire State Building and all constructions in Europe, Africa, and Latin America (Figure \ref{TheLine}).

\begin{figure}[!htbp]
\centering
\includegraphics[width=0.7\textwidth]{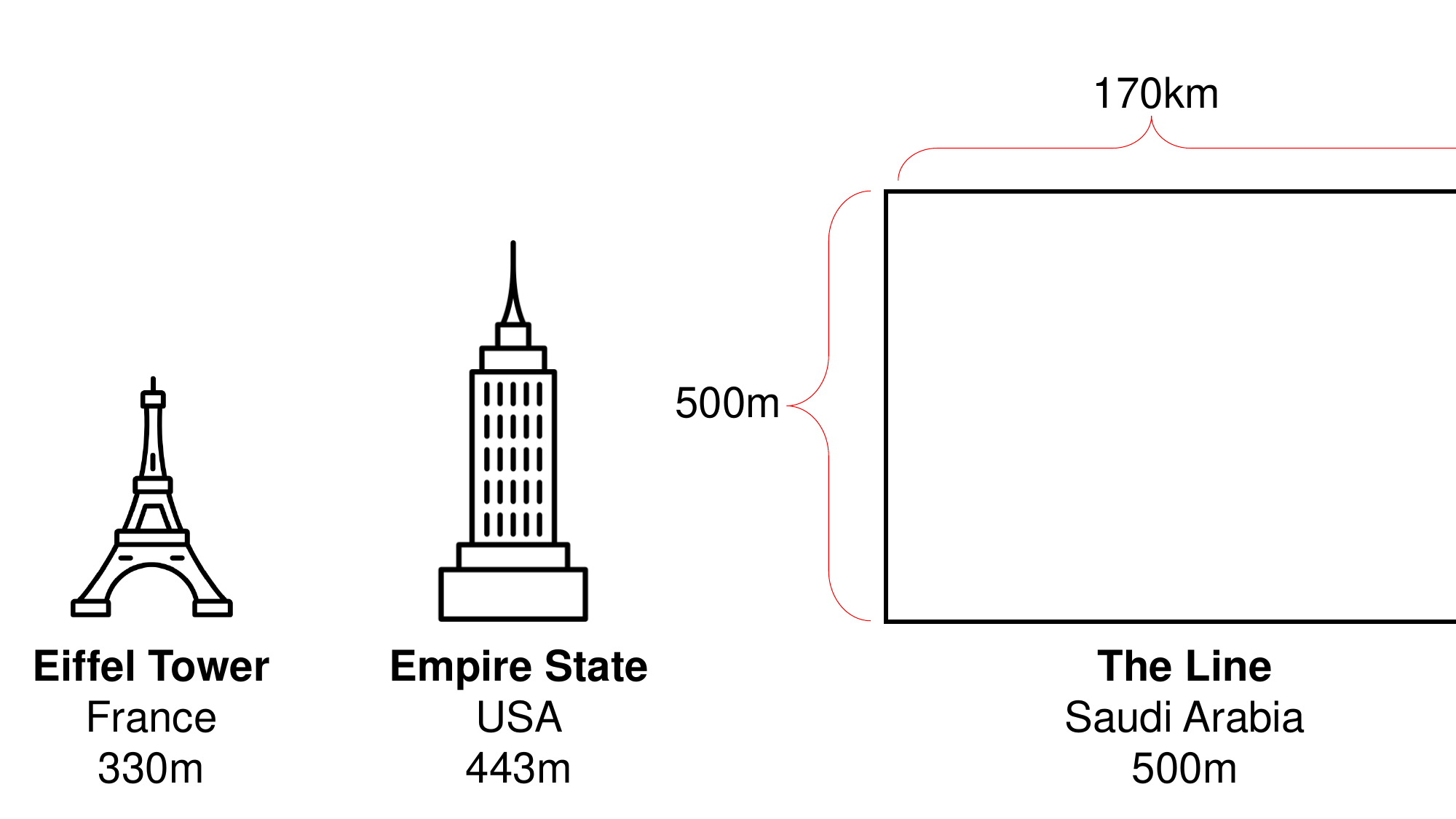}
\caption{The Line is a planned city in Saudi Arabia. Its plan includes an elongated infrastructure of 170 km long, spanning a 500 m tall continuous building.}\label{TheLine}
\end{figure}

}

{
In The Line, we can consider the whole concept as a single building (so $B_L = 1$ building), with a surface of 34 km\textsuperscript{2} (so $A_L = 34$ km\textsuperscript{2}), with a maximum distance of 170 km from one extreme to the other (so $M_L = 170$ km). The Elongation of The Line is
\begin{equation}
E_L = \frac{\sqrt{\pi} M_L}{2\sqrt{B_L A_L}} = 25.8,
\end{equation}
which is an extremely large value of its Elongation. Further, the average distance between two random locations in a line of length $l$ is $l/3$, so we get that $D_L = 170/3$ km. Thus, we get that the Sprawl of The Line will be
\begin{equation} 
S_i = \frac{45^2 \pi^{3/2} D_L^2}{2^{13} M_L \sqrt{B_L A_L}} = 4.5,
\end{equation}
also highly sprawled. Therefore, in The Line, their extreme Elongation and Sprawl contribute to the distance between people in the city being $\sqrt{S_L E_L} \approx 10.7$ times longer than in a circular city \cite{prieto2023arguments}. 
}

{
Although the shape of The Line has some negative aspects, including its extreme elongation and the need for extremely tall, costly infrastructure to achieve such high densities, there are at least two positive aspects for the city. First, they aim to minimise the footprint of the city. Thus, the plans for The Line start by recognising that urban sprawl has negative impacts on the planet. But secondly, they aim to construct a car-free city. Thus, they also recognise that modern urban mobility should rely primarily on active mobility for short distances and on public transport for longer journeys.
}

\subsection{Urban scaling and ownership}

{
We can analyse ownership at the city level using the principles of urban scaling. Consider, for example, how many cars are in a city. If the probability that a person owns a car does not vary with the size of the city they live in, then we should observe a fixed ratio between the number of people and the number of cars (for example, one car for every four people). Yet if city size affects how we move, it could also affect the number of cars we own. Considering 75 metropolitan areas in Mexico shows that, on average, there are roughly 15 domestic cars per 100 people (including vans). Yet, the data also reveals some scaling correlations between car ownership and city size (Figure \ref{ScalingOwnership}). 

\begin{figure}[!htbp]
\centering
\includegraphics[width=0.99\textwidth]{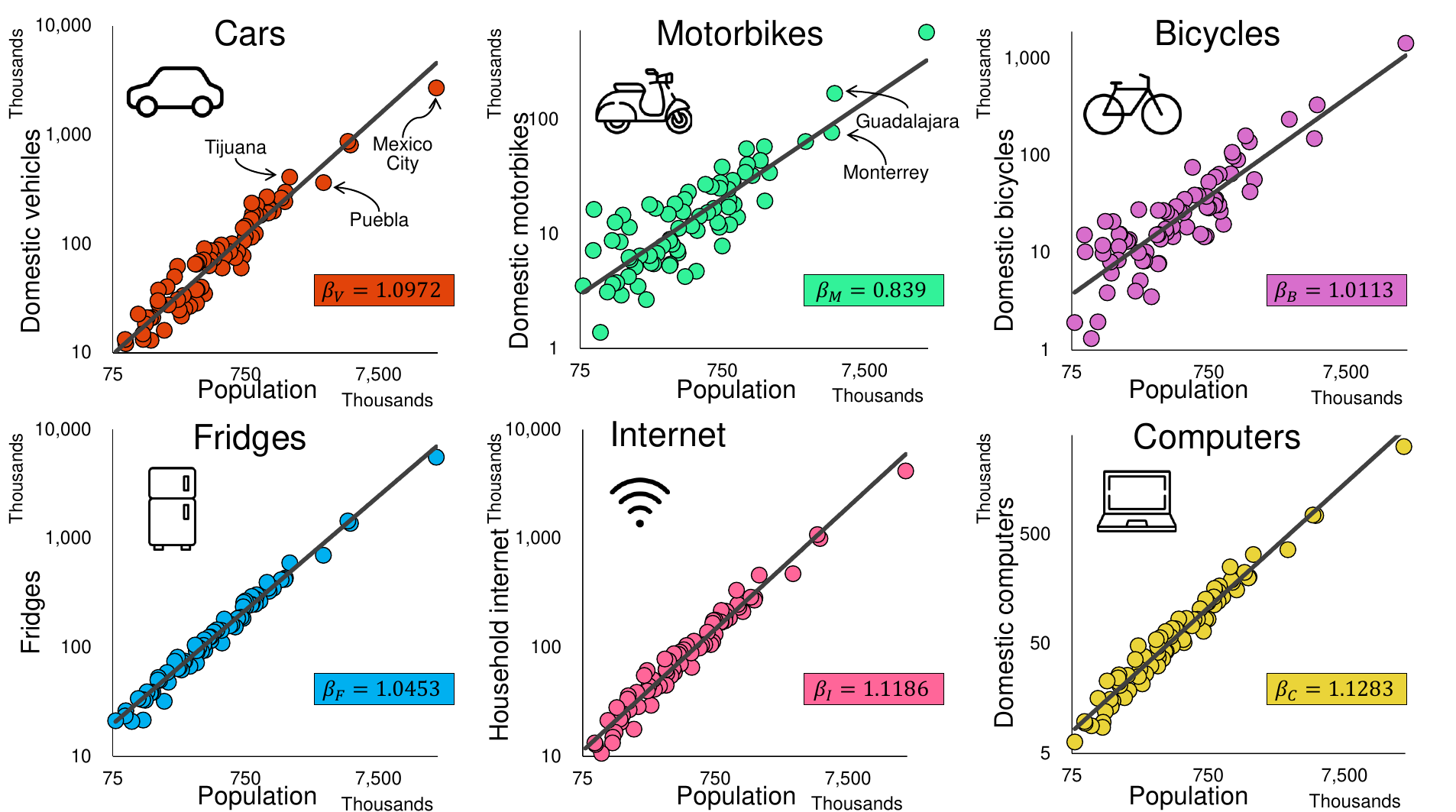}
\caption{Number of cars and vans (vertical axis), the number of motorbikes, the number of bicycles, fridges, households with internet and computers in 75 metropolitan areas in Mexico, according to the population of each city (horizontal axis). Both axes are represented on a logarithmic scale. Data from the 2020 census \cite{Inegi2020}. }\label{ScalingOwnership}
\end{figure}
}

{
We estimate the scaling of the number of cars in city $i$ by expressing
\begin{equation}
V_i = \alpha_V P_i^{\beta_V},
\end{equation}
for some values of $\alpha_V$ and $\beta_V$. We obtain that $\alpha_V = 0.0434 \pm 0.0259$ and that $\beta_V = 1.0972 \pm 0.0359$. Since $\beta_V$ is slightly greater than one, in large cities, there is a higher number of cars per person. If we compare two cities in Mexico, $C_1$ and $C_2$, where the population of $C_2$ is two times the population of $C_1$, the expected number of cars in $C_2$ is approximately $2^{1.0972} = 2.14$ times larger. Therefore, in city $C_2$, the probability that a person owns a car is 7\% higher than in city $C_1$. 
}

{
There is a similar pattern for the ownership of motorbikes and bicycles. In Mexico, there are 2.6 motorbikes per 100 people, but the rate varies across the country. We find that for motorbikes, $\alpha_M = 0.2402 \pm 0.2976$ and $\beta_M = 0.8389 \pm 0.0618$, so there are fewer motorbikes per person in larger cities. Also, in Mexico, there are 5.7 bicycles for every 100 people, and we find that $\alpha_B = 0.0437 \pm 0.0627$ and $\beta_B = 1.0113 \pm 0.0683$, meaning that if we consider the number of bicycles per 100,000 inhabitants, we find a similar number in small or big cities (Figure \ref{ScalingOwnership}). City size influences mobility choices, the feasibility of different modes of transport, and even the ownership of cars and motorbikes \cite{MobilityABCPrietoOspina}. 
}

{
It has been observed that, for cities outside the US, size has a significant influence on transport patterns \cite{MobilityABCPrietoOspina}. In small cities outside the US, Active mobility (including walking, cycling, skating, and others) and Car journeys are more common, but public transport options are limited, and for large cities, there is a shift towards public transport as the primary mode of transportation \cite{MobilityABCPrietoOspina}.
}

{
We can also analyse the impact of city size and ownership of other home appliances. For example, in Mexico, there are roughly 27 fridges per 100 people, but the rate varies by city size. We observe that for the number of Fridges, $\alpha_F = 0.1589 \pm 0.0376$ and $\beta_F = 1.0452 \pm 0.0163$, meaning that in larger cities, there is slightly higher ownership of a Fridge (Figure \ref{ScalingOwnership}). Similarly, for households with access to the Internet, $\alpha_I = 0.0395 \pm 0.0151$ and $\beta_I = 1.1186 \pm 0.0248$, meaning more access in larger cities. Finally, for the number of domestic Computers, we find that $\alpha_C = 0.0251 \pm 0.0102$ and $\beta_C = 1.1282 \pm 0.0263$, so there is also a higher number of computers for domestic use in large cities.  
}

{
Again, if we compare two cities, $C_1$ and $C_2$, where the population of $C_2$ is two times the population of $C_1$, the expected number of Fridges per person is 3\% higher in $C_2$. Also, in the city $C_2$, there is a 9\% higher probability of having domestic access to the Internet and a 9\% higher probability of owning a Computer for domestic use (Figure \ref{ScalingOwnership}). 
}

{
Some results from urban scaling studies need to be interpreted with caution. For example, we observed the number of vehicles in cities of varying size (Figure \ref{ScalingOwnership}). We observe that, since $\beta_V > 1$, there are more cars per person in large cities, and Figure \ref{ScalingOwnership} somewhat confirms this scaling behaviour. Yet, both axes are displayed on a logarithmic scale, so the ``small'' error the model has for the largest city in the country (Mexico City) appears to be of similar magnitude to the error observed across other cities. However, if we observe the same model without the logarithmic scale, the errors produced for that city become more evident (Figure \ref{ScalingErrors}). The model estimates 4.6 million cars in Mexico City, 1.9 million above the observed number. Thus, the model overestimates the number of cars by a whopping 69\%. Using a logarithmic transformation reduces errors that are systematically produced by scaling models.

\begin{figure}[!htbp]
\centering
\includegraphics[width=0.99\textwidth]{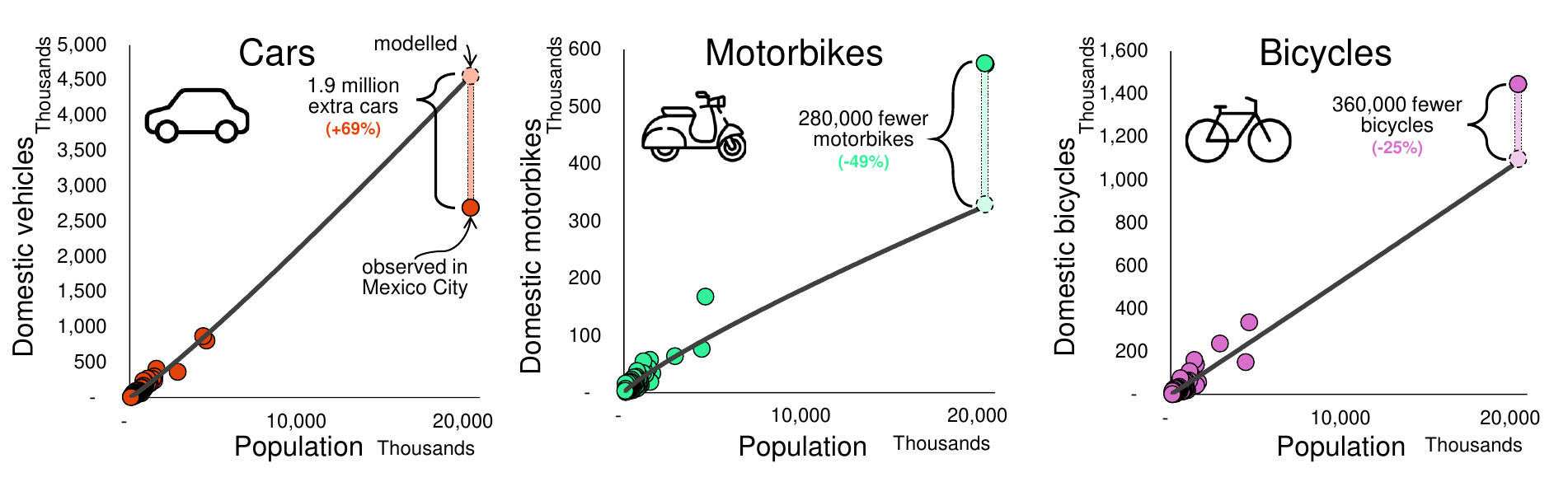}
\caption{Number of cars and vans (vertical axis), the number of motorbikes and the number of bicycles in 75 metropolitan areas in Mexico according to the population of each city (horizontal axis).}\label{ScalingErrors}
\end{figure}
}

{
The same type of error occurs for the number of motorbikes and bicycles in the country. According to the scaling model, there should be 49\% fewer motorbikes in Mexico City and 25\% fewer bicycles (Figure \ref{ScalingErrors}). Thus, it is relevant to consider that some aspects of a scaling model might result from a visual display, in which the results may be perceived as obvious. 
}

{
If instead of analysing the raw number of cars in a city, we look at the number of cars per 100 people, the results become less evident (Figure \ref{ScalingPer100}). Now, the number of domestic vehicles per 100 people does not seem to have an obvious relationship with city size, and the same holds for the number of motorbikes and bicycles per 100 people.

\begin{figure}[!htbp]
\centering
\includegraphics[width=0.99\textwidth]{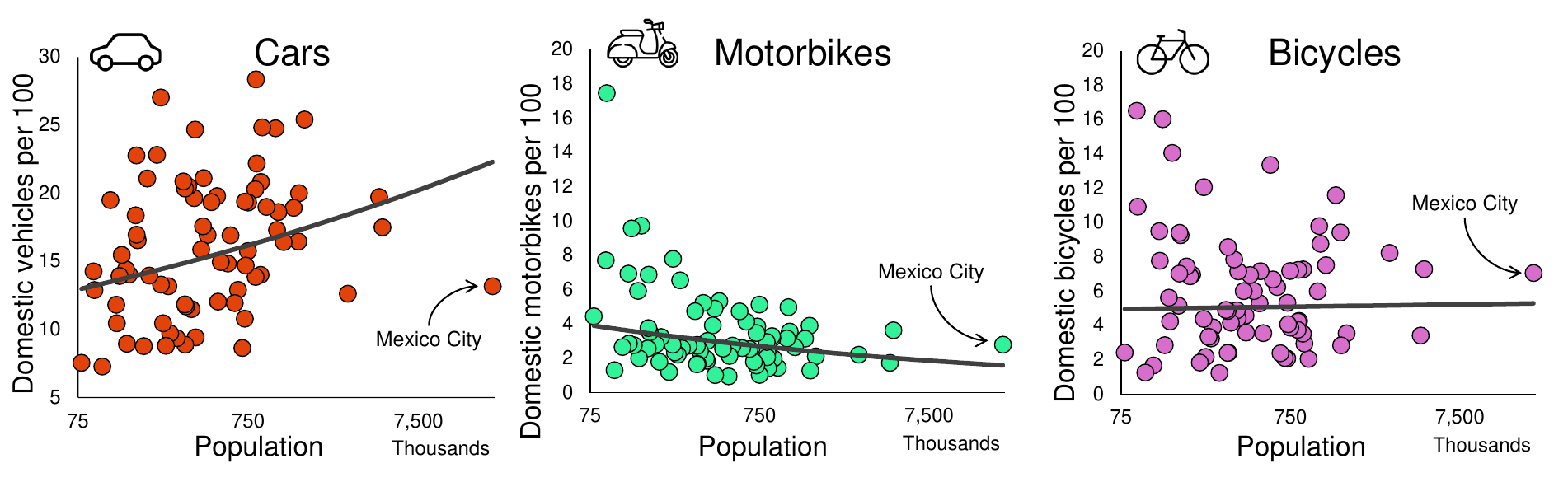}
\caption{Number of cars and vans (vertical axis), the number of motorbikes and the number of bicycles per 100 people in 75 metropolitan areas in Mexico according to the population of each city (horizontal axis). The horizontal axis (population) is displayed on a logarithmic scale.}\label{ScalingPer100}
\end{figure}
}

{
Finally, there is also another issue related to the observations considered. In the case of the 75 metropolitan areas of Mexico, we assign the same weight to information from Mexico City (a metropolitan area with 21 million inhabitants) as to information from Rioverde (a metropolitan area with fewer than 100,000 inhabitants). In fact, the metropolitan area of Mexico City has a larger population than the smallest 57 metropolitan areas combined. Should the results assign the same weight to all observations? One option is to consider a weighted regression, where greater weight is given to cities with larger populations, assuming these observations provide more information. When we consider a weighted regression to estimate the scaling coefficients, we obtain $\beta_V^{(w)} = 0.9582 \pm 0.0178$, indicating that, within this scope, the number of vehicles per city is sublinear. Similarly, the number of motorbikes is linear, with  $\beta_M^{(w)} = 0.9951 \pm 0.0274$, and the number of bicycles is superlinear, with $\beta_B^{(w)} = 1.0897 \pm 0.0310$.
}

{
Estimating the scaling coefficients is challenging. Under different methods, the observed variable can be considered either superlinear (as in the case of cars in cities in Mexico) or sublinear (as in the case of cars in cities in Mexico, but when coefficients are estimated using a weighted regression). Therefore, estimating the scaling parameters can yield conflicting results. However, scaling studies also face other challenges beyond parameter estimation. Similar issues have been observed elsewhere, where some scaling even changes the sublinearity to superlinearity \cite{cottineau2017diverse}.

}

\subsection{The challenges of urban scaling studies}

{
Urban scaling studies have many challenges. Firstly, defining urban agglomerations is difficult and depends on considerations and parameters \cite{arcaute2015constructing, rozenblat2020extending}. What a city actually is and how to compare them are not at all trivial matters. Urban scaling values are not universal and depend significantly on how one constructs cities and on the variables considered \cite{cottineau2017diverse}. The way cities or metropolitan areas are defined may alter results \cite{arcaute2013city, oliveira2014large, rozenfeld2008laws}. This issue is particularly relevant when cities from different countries are compared. Particularly in some developing regions, census data might be outdated and have critical implications. For example, the latest available census in Somalia was carried out decades ago, but in Kenya, a neighbouring country, the most recent census was in 2019. Thus, comparing cities in East Africa may be challenging.
}

{
One of the biggest challenges in urban scaling studies is that many of the results obtained so far are based on observations from the USA. However, more data from other regions of the world have opened new studies that challenge the universality of scaling values. For example, observing the shapes of 5,000 cities in Africa revealed that their surface area is linearly correlated with city size \cite{prieto2023scaling}. However, a non-trivial growth in the average building size, with an approximately linear increase in the number of buildings and in urban form indicators, increases the expected distance between different parts of the city \cite{prieto2023scaling}.
}

{
Yet new technology is also opening up new ways to observe populations and cities. For example, satellite images are now a valuable tool for observing population settlements and estimating the number of inhabitants at a very refined scale. Thus, this type of technology is often provides more accurate information than a census that was carried out decades ago. Using this type of input, a novel dataset that delineates all African cities is now available. Africapolis maps the location and boundaries of more than 5,000 urban agglomerations based on the same definition for an urban agglomeration at a continental level \cite{Africapolis}. It is based on a city's infrastructure and, using aerial images, delineates a city if its buildings are less than 200 m apart. Thus, scaling studies outside the USA are now more feasible.
}

{
Also, beyond numerous urban scaling correlations, it is crucial to understand the driving forces and mechanisms behind them. For example, it is possible to measure greenhouse gas emissions at the city level and find that large cities emit less per person than smaller cities \cite{lu2024worldwide}. Yet, although the result is relevant by itself, it is unclear why there is a reduction in larger cities. Do people in large cities make fewer trips on a regular week than people in small cities? Are there fewer emissions because people are more likely to use Public Transport? Are people less likely to drive in big cities, or is it related to infrastructure, such as the increase in electric cars? Or are bigger cities producing fewer greenhouse gas emissions because they outsource most of their industry to other places? Without an understanding of the mechanisms underlying the observed correlations, little is known about their implications or which policies should be implemented to improve urban systems. If large cities produce fewer emissions only because they outsource industrial activities somewhere else, the repercussions are critically different than a reduction related to mobility \cite{MobilityABCPrietoOspina}. 
}

{
There are many ways to understand cities beyond correlational studies. For example, cellular automata and agent-based models, combined with machine learning techniques, have become powerful tools for simulating complex processes within urban dynamics \cite{CitiesComplexity, duque2019spatiotemporal}. How will a city expand due to population growth, enabling the provision of services such as electricity and water infrastructure \cite{duque2019urban}. These type of models usually consider the complex interactions between agents (say, urbanites), their neighbourhood and its infrastructure. Although agent-based models face challenges related to calibration and validation, they enable consideration of the main mechanisms and driving forces within urban dynamics \cite{heppenstall2020future}.
}

\section{Urban studies beyond scaling} 

{
Urban scaling studies tend to analyse infrastructure or social aspects of a city in terms of a single indicator, for example, in terms of income, accessibility, road insecurity or crime. Yet for most urban indicators, there is substantial heterogeneity within a single city. For example, in terms of proximity to public transport, central areas of a city tend to be better connected and enjoy more provisions than remote areas. There is also an issue with the spatial arrangement of the city's areas. In theory, even if a metropolitan area is large, people could still live close to their destinations. However, a neighbourhood is not a small-scale version of a city. Consider, for example, the number of crimes in a city. Take, for example, the crimes reported to the Police in Mexico City (Figure \ref{CrimeIntensity}). Most neighbourhoods have a relatively low level of crime, and only a few locations in the city tend to have the highest crime rates.

\begin{figure}[!htbp]
\centering
\includegraphics[width=0.7\textwidth]{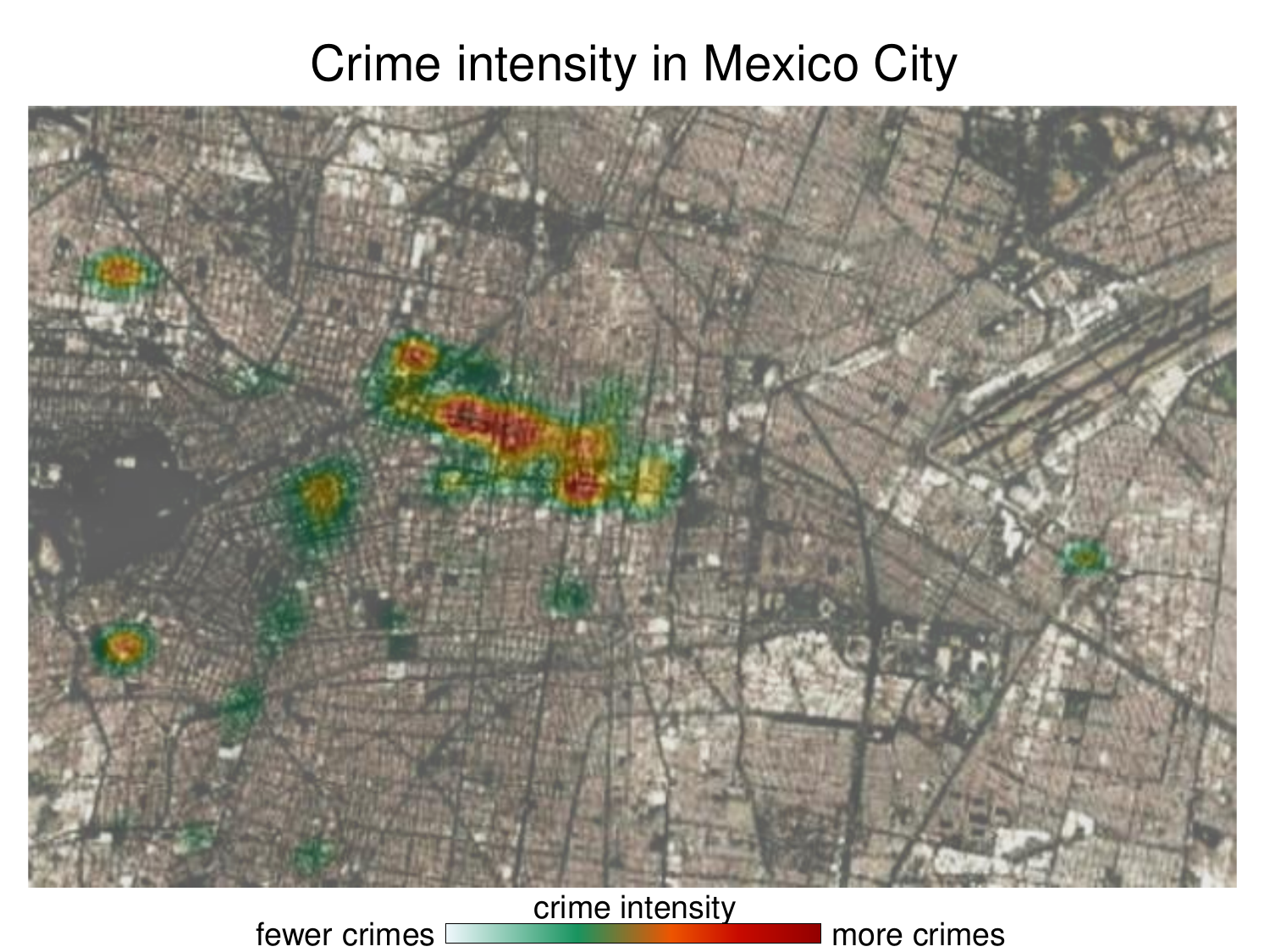}
\caption{Intensity of crimes reported to the Police in Mexico City between 2016 and 2020. It includes only property crimes (robbery of a person, car theft, shoplifting, and burglary). The intensity is computed using a 2-D kernel density estimate based on the locations of reported crimes. Neighbourhoods with fewer crimes have transparent backgrounds, while areas with more crimes are shaded in green, orange, and red, in that order.}\label{CrimeIntensity}
\end{figure}
}

{
Crime is highly concentrated among different neighbourhoods or street segments, resulting in hotspots of crime \cite{LawCrimeConcentration, SystematicReviewPlaces}. This pattern is roughly stable in time (so if we compare the hotspots from 2016, they are roughly the same as the ones in 2019) and tends to reveal areas with the highest social activities, like work or schools, transport stations, and more. Thus, we see that crime is highly heterogeneous at the city level. In fact, similar to what we observe at the neighbourhood level, we see it at the personal level. It was observed that roughly 60\% of the population in Mexico City is ``immune'' to crime, meaning they will rarely experience it \cite{ConcentrationOfCrimeRECC}. However, in Mexico City, some people (perhaps less than 5\% of the population) are chronically victimised, meaning that they tend to suffer crime frequently \cite{ConcentrationOfCrimeRECC}. Therefore, perhaps the number of crimes per person is not a relevant indicator if some people within the same city are frequently victimised, whilst most people are rarely the victims of crime.
}

{
Analysing the correlation between city size and crime rates is relevant, as it helps us understand whether people in large cities tend to experience higher or lower crime rates than those in smaller cities. However, analysing the spatial distribution at different spatial scales within a city is also a crucial part of understanding urban dynamics, for example, by comparing different neighbourhoods in a city, street segments, or at the individual level. 
}

{
Although there are many challenges to analysing cities at smaller spatial units (including the delineation of neighbourhoods and how arbitrary or political boundaries may strongly affect the results), a relevant aspect to consider is the extent to which a variable is autocorrelated across space. For example, a neighbourhood with a high crime rate is likely to be close to other neighbourhoods with high crime rates. Furthermore, a neighbourhood's position within a city might also be a critical factor. Some parts of a city tend to attract workers or students from other areas, whilst other parts, mainly the remote suburbs, may lose large parts of their population due to daily routines \cite{felson2015daily}. In most cities, a small fraction of the neighbourhoods tend to monopolise most workers, students, shoppers, and visitors. 
}

\subsection{Comparing at a smaller scale between cities}

{
One way to preserve parts of a city's geography is to consider its position relative to its centre. For a city, we first identify its centre. This could be, for example, the location of the main square, temple, or transport station. Although the precise location of the centre could be ambiguous (for example, if there are many candidates, such as a central square or transport station), we pick one of those locations that may be used to determine the relative proximity of other places in the city. Based on the city's centre location, we measure the distance to it. Let $D_{i,j}$ be the distance in km for location $j$ in city $i$ to its centre. With this metric, remote locations are identified by having a large $D_{i,j}$. Although the city centre could be defined in terms of more than one location, the impact on the distance to the city centre is usually small. In Paris, France, for example, the distance from Stade de France (a stadium in the northern part of the city) is 7.3 km from the Louvre, 7.8 km away from Notre-Dame, and 7.2 km from Place de la Concorde (three candidates to be considered the city centre). 
}

{
After we identify the centre of a city, we can quantify an urban indicator at the neighbourhood level, for example, the number of car crashes, the number of crimes, or some economic indicator, such as the number of households with access to water, a car, or a computer. Then, groups of different locations that share their distance from the centre can be formed. This way, we analyse concentric rings covering the whole city. The idea behind this is that central locations in a city differ from intermediate or remote locations. For example, consider the percentage of households that own a computer. Census data from Mexico in 2020 enable us to observe ownership of certain items at the census-tract level (Figure \ref{DistRemoteness}). Although there is a huge variability in terms of the number of households with a computer, it is possible to notice a pattern across all cities. In central areas, owning a computer is more common (between 40 and 80\% of households have a computer). Yet, for remote locations, owning a computer is much less frequent. If the distance to the centre is above a certain threshold, then owning a computer is much less frequent (between 20 and 40\% if the distance is above 30 km).

\begin{figure}[!htbp]
\centering
\includegraphics[width=0.7\textwidth]{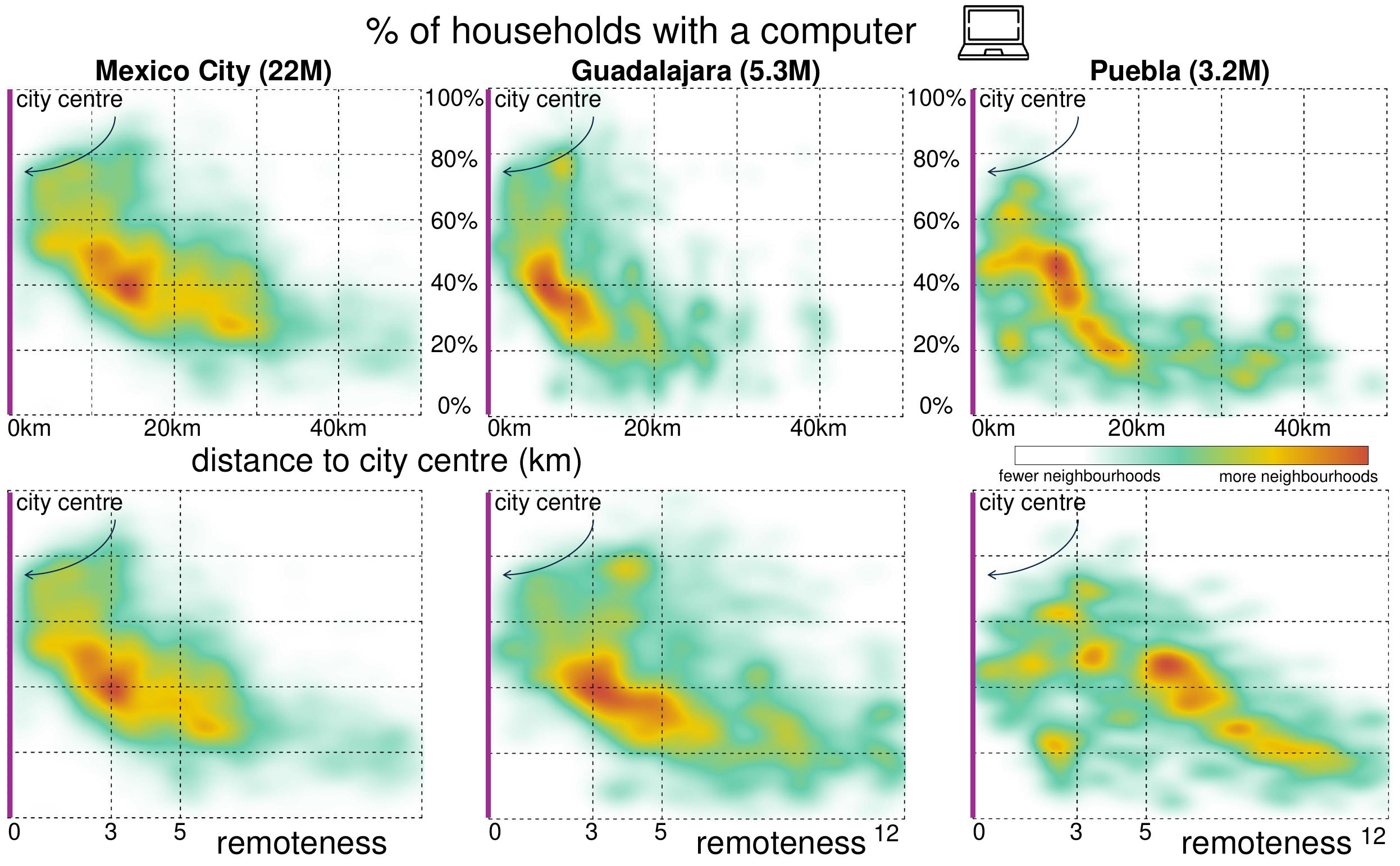}
\caption{Taking census tracts in three cities in Mexico, we measure the number of households that own a computer (vertical axis). The centre is located at the left part of each figure, and the distance (horizontal axis) indicates all census tracts that are up to 50 km away from the centre. The figures correspond to Mexico City (left), Guadalajara (centre), and Puebla (right). The top panels show the distance to the centre, and the bottom panels show the remoteness.}\label{DistRemoteness}
\end{figure}
}

{
The distance to a city centre, however, is relative to the population of a city. For example, in Mexico City (a metropolitan area with 22 million inhabitants), a location 2.5 km from the city centre is still considered central. However, in Puebla (a city with 3.2 million inhabitants), a location within the same distance is already distant. And in even smaller cities, 2.5 km away from the city centre would already be considered outside of its boundary. One way to remove the effect of size is as follows. Imagine a city with a population $P_i$, where each person occupies the same area, say $\nu > 0$ m\textsuperscript{2}. Then, its area can be expressed as $\nu P_i$. If that city is arranged circularly and compactly, then we can compute the average distance between two people. If we take two random points inside that circle, the expected distance between them is
\begin{equation}
\frac{128}{45 \pi} \sqrt{ \nu P_i} = \delta \sqrt{P_i},
\end{equation}
for $\delta = 128 \sqrt{\nu}/45 \pi$, meaning that distances in a city grow with the square root of its population \cite{prieto2023scaling}. Thus, by considering the location $j$ in city $i$ and dividing by the square root of its population, we essentially remove the impact of city size \cite{WaterPrietoBorjaArxiv}. We construct the \emph{remoteness} \index{Remoteness} of location $j$ in city $i$, expressed as $R_{i,j}$ by
\begin{equation} \label{Rem}
R_{i,j} = \frac{1000 D_{i,j}}{\sqrt{P_i}}.
\end{equation}
The expression \ref{Rem} is multiplied by 1000 to obtain numbers that are easier to manipulate. With this formula, a location in Mexico City that is 2.5 km away from its centre has $R_{X,j} = 0.5$, whereas a location in Puebla 2.5 km away from its centre has $R_{P,j} = 1.4$, so the same distance would be considered much more remote. In a small city $S$ with only $P_S = 100,000$ inhabitants, 2.5 km away from its centre has $R_{S,j} = 7.9$. 
}

{
The concept of remoteness provides a way to understand cities, revealing how urban amenities tend to concentrate in central areas, leaving remote regions with fewer amenities and higher levels of social exclusion. Remote areas generally experience higher levels of poverty and social exclusion. For example, looking at the percentage of households that own a computer, there are many more similarities across cities (Figure \ref{DistRemoteness}). For a small level of remoteness (for example, with $R_{i,j}<5$), there is a wide heterogeneity in terms of owning a computer. Many regions in a city's central areas are wealthier, and people tend to own computers. However, in remote areas (for example, with $R_{i,j}>5$), neighbourhoods tend to be poorer, so there are fewer households with a computer. 
}

\section{Conclusions}

While urban scaling theories offer insights into how urban metrics change with population size, they are not universal laws for all cities. Rather, they provide a framework to analyse cities of varying sizes. Results provided by urban scaling studies need to be carefully analysed since they depend on many aspects, including the definition of cities and the parameters used. Further, if in some region it has been observed that large cities tend to share certain attributes, this might not be a permanent pattern, and there is no reason to assume it will apply to cities in other regions. 

Scaling studies overlook significant heterogeneity within cities. Factors such as spatial arrangement, proximity to amenities, and socio-economic disparities play crucial roles in shaping different neighbourhoods within the same city. Analysing cities at finer spatial scales, such as neighbourhood levels, reveals essential patterns of variation. We might find clusters of similar characteristics, emphasising the interconnectedness of neighbouring areas. Moreover, considering the distance from a neighbourhood to the city centre helps capture gradients of infrastructure and socio-economic indicators, highlighting disparities between central and remote locations. This situation underscores the importance of analysing urban dynamics beyond scaling frameworks, considering spatial heterogeneity, and exploring local variations within cities. Such approaches are essential for developing targeted policies and interventions to address the diverse needs and challenges faced by different neighbourhoods within urban areas.

\bibliographystyle{unsrt}

\end{document}